\documentclass[pra,amsmath,amssymb,aps,showpacs,10pt,superscriptaddress,twocolumn,floatfix]{revtex4-1}
\usepackage[utf8]{inputenc}
\usepackage{amsmath}
\usepackage{latexsym}
\usepackage{amsfonts}
\usepackage{epsfig}
\usepackage{psfrag}
\usepackage{float}
\usepackage{textcomp}
\usepackage{gensymb}
\usepackage{natbib}
\usepackage{graphicx}
\usepackage{skmath}
\usepackage{comment}
\usepackage{siunitx} 
\usepackage{upgreek}
\newcommand\dscript[2]{\bgroup#1\egroup_\bgroup#2\egroup}

\begin{document}
\title{Measurement of electric-field noise from interchangeable samples with a trapped-ion sensor}
\author{Kyle S. McKay}
\affiliation{National Institute of Standards and Technology, 325 Broadway, Boulder, Colorado 80305, USA}
\affiliation{Department of Physics, University of Colorado, Boulder, Colorado 80309, USA}

\author{Dustin A. Hite}
\affiliation{National Institute of Standards and Technology, 325 Broadway, Boulder, Colorado 80305, USA}

\author{Philip D. Kent}
\affiliation{National Institute of Standards and Technology, 325 Broadway, Boulder, Colorado 80305, USA}
\affiliation{Department of Physics, University of Colorado, Boulder, Colorado 80309, USA}

\author{Shlomi Kotler}
\affiliation{Department of Applied Physics, The Hebrew University of Jerusalem, Jerusalem 9190401, Israel}

\author{Dietrich Leibfried}
\affiliation{National Institute of Standards and Technology, 325 Broadway, Boulder, Colorado 80305, USA}

\author{Daniel H. Slichter}
\affiliation{National Institute of Standards and Technology, 325 Broadway, Boulder, Colorado 80305, USA}

\author{Andrew C. Wilson}
\affiliation{National Institute of Standards and Technology, 325 Broadway, Boulder, Colorado 80305, USA}

\author{David P. Pappas}
\affiliation{National Institute of Standards and Technology, 325 Broadway, Boulder, Colorado 80305, USA}

\date{\today}
\begin{abstract}
We demonstrate the use of a single trapped ion as a sensor to probe electric-field noise from interchangeable test surfaces. As proof of principle, we measure the magnitude and distance dependence of electric-field noise from two ion-trap-like samples with patterned Au electrodes. This trapped-ion sensor could be combined with other surface characterization tools to help elucidate the mechanisms that give rise to electric-field noise from ion-trap surfaces. Such noise presents a significant hurdle for performing large-scale trapped-ion quantum computations. 
\end{abstract}
\maketitle

\section{Introduction}
 Electric-field noise in ion traps can result in heating or dephasing of ion motion, causing reduced gate fidelities in an ion-trap quantum computer \cite{Wineland1998}. Measured heating rates in trapped-ion experiments are typically orders of magnitude higher than the heating expected from thermal (Johnson) noise and known technical noise sources. For this reason, the excess heating has been termed `anomalous' by the ion-trap community. Anomalous heating has been attributed to electric-field noise that emanates from the surfaces of ion-trap electrodes, in large part due to the apparent scaling of the noise as $\sim d^{-4}$, where $d$ is the distance between the ion and the nearest electrode \cite{Wineland1998, Turchette2000, Deslauriers2006, Brownnutt2015}. A $d^{-4}$ scaling is expected in the case of an infinite surface covered with independent fluctuating patch potentials, where the radius of the patches is much smaller than $d$. Fluctuations of the potential of an entire electrode, as would be the case for Johnson noise, results in noise that scales as $d^{-2}$ for electrodes with dimensions $\gtrsim d$ \cite{Turchette2000}. In recent years, further evidence for surface origins of the noise has been provided by experiments that varied the bulk resistivity \cite{Wang2010} and temperature \cite{Chiaverini2014} of trap electrodes and by surface treatments \cite{Hite2012, Daniilidis2014, McKay2014, Sedlacek2018a} that reduced electric-field noise by as much as two orders of magnitude as compared to electric-field noise from untreated electrode surfaces. Despite this improvement, the underlying physical mechanisms that result in anomalous heating are not understood. 
 
In Ref. \cite{Hite2017}, a surface-science system was combined with a stylus ion trap to enable $in$ $situ$ preparation and characterization of a sample surface combined with electric-field-noise measurements of the surface. This system was designed to investigate correlations between the electric-field noise from surfaces and surface properties such as composition and morphology. However, in that system, electric-field noise from the samples was not detectable over the background electric-field noise from the ion-trap electrode surfaces. A hypothesis that anomalous heating might arise from rf driven surface processes on ion trap electrodes was presented as a possible explanation for the inability to detect heating from these samples. Here we revisit this experiment with changes to the sample electrode patterning and electrical connections, and present evidence that anomalous heating due to the sample can be measured by an ion in the stylus-trap system. We also explore the hypothesis that rf driving may be correlated with an increase in electric-field noise from the sample surface. 

\section{Experimental Setup}
\subsection{Stylus Trap}
The ion trap in this work (see Fig. 1) has a stylus geometry \cite{Maiwald2009, Arrington2013} to enable samples to be positioned close to the ion while still allowing for appropriate laser access to and fluorescence collection from the ion. 
\begin{figure*}
\centering
\includegraphics[width=0.9 \textwidth,keepaspectratio]{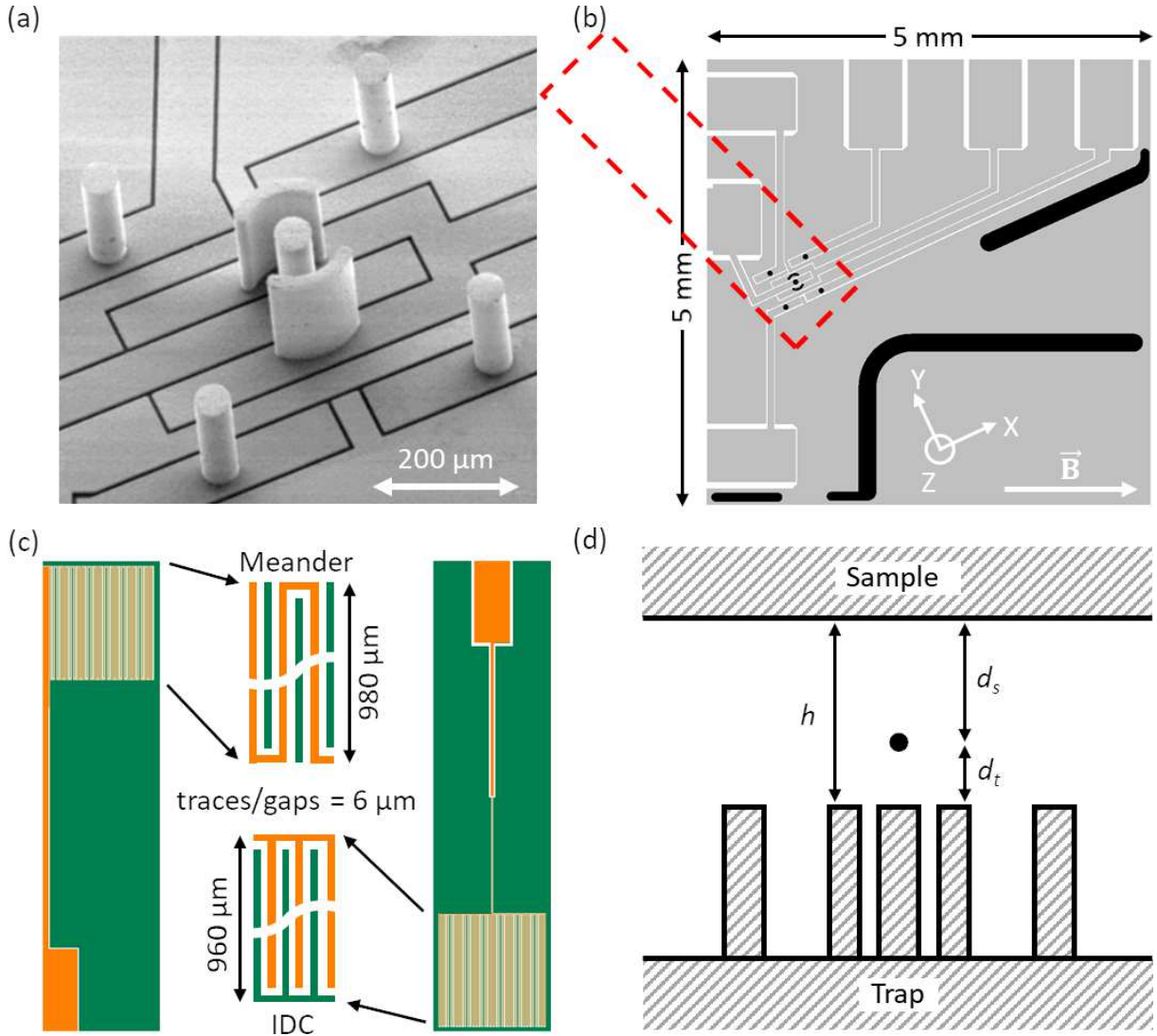}
\caption{a) Scanning electron microscope image of an example stylus trap (similar to the trap used in this work). The trap consists of a pair of symmetric arcs that form the rf electrode, along with one central and four surrounding cylindrical dc electrodes. The \SI{60}{\um}-diameter cylindrical posts and the symmetric arcs are extruded to a height of $\sim$\SI{180}{\um}. b) Top view layout of the $5$ mm $\times$ $5$ mm stylus trap chip. The positioning of the sample chips relative to stylus features is shown by the dashed red outline. The normal-mode directions for a single ion are approximately overlapped with the Cartesian axes shown and the static magnetic field is oriented at $25^\circ$ and $86.5^\circ$ relative to the $X$- and $Z$-axis, respectively. c) Schematics of the meander and interdigitated capacitor (IDC) samples along with diagrams of the pattern of the fine featured electrodes in the sample regions. The signal electrode is shown in orange and the grounded electrode is shown in green. The width of the gaps and traces within the $\sim$\SI{1}{mm} $\times$ \SI{1}{mm} area of interest opposite the large bond pads is \SI{6}{\um}. d) Schematic showing the distances involved in the heating-rate measurement. The sample distance $h$ relative to the stylus electrode top plane is equal to the sum of the ion-to-trap distance, $d_{t}$, and the ion-to-sample distance, $d_{s}$. The view shown in the schematic is along the $X$-axis. The schematic is not drawn to scale.}
\label{fig:Trap}
\end{figure*}
The trap consists of a pair of symmetric vertically extruded arc-shaped electrodes to provide rf confinement of an ion, along with one central and four surrounding cylindrical dc electrodes (see Fig. 1). The cylindrical posts, used to compensate for stray electric fields, have a diameter of \SI{60}{\um}. The rf electrode arcs both have a central angle of $84 \degree$ and inner and outer radii of \SI{72}{\um} and \SI{98}{\um}, respectively. The rf arcs and the compensation posts are composed of electroplated Au and are extruded to a height of \SI{180}{\um} above the underlying electrical layer. This layer consists of \SI{5}{\um}-thick electroplated Au on a quartz substrate. Details of the fabrication process can be found in Ref. \cite{Arrington2013}. 

In the absence of a sample, the ion height above the stylus electrodes, $d_{t0}$ [see Fig. 1(d)], is determined to be $\sim 59\pm$\SI{2}{\um} by scanning a laser beam with \SI{20}{\um} diameter at 1/e$^2$ intensity across the trap features and/or the ion using a servo motor with sub-micrometer step size. The trap electrode position is determined by observing the obscuration of the transmitted beam by the trap electrodes as a function of the beam's position. The ion position is determined by measuring the position where this laser beam, when detuned from a resonant atomic transition, induces the maximal ac Stark shift on this transition. The pseudopotential produced by the rf electrodes gives rise to normal-mode directions for a single ion that are parallel ($x$) and perpendicular ($y$) to the gap between the rf electrodes and perpendicular to the trap substrate ($z$). The $x$, $y$, and $z$ normal-mode directions are approximately overlapped with the $X$, $Y$, and $Z$ Cartesian axes shown in Fig. 1(b). The trap frequency $\omega$ ratios are $(\omega_x, \omega_y, \omega_z)/\omega_z=(0.27, 0.73, 1)$. When a sample is positioned above the trap [along the $Z$-axis, see Fig. 1(d)], the sample acts as an additional electrode and affects the curvatures and the position of the pseudopotential minimum. As the trap-to-sample distance $h$ is reduced, the distance $d_t$ between the pseudopotential minimum (where the ion is trapped) and the stylus trap electrodes is reduced. Additionally, the motional mode frequencies increase and the difference in $\omega_x$ and $\omega_y$ ratios decreases. The change in ion height $\Delta d_{t}$  is measured by monitoring the position of the ion imaged with an objective onto an electron-multiplied charge-coupled device (EMCCD) camera. The effective resolution of the imaging system is $13.5$ pixels per micrometer. The ion-to-trap distance $d_t$ is given by $d_t = d_{t0} - \Delta d_t$, where $\Delta d_t$ is a function of the trap-to-sample distance $h$ [see Fig. 2(b)]. 

The rf confining potential is provided by an rf synthesizer and a helical resonator with a loaded quality factor of 160 at 64 MHz to step up the drive voltage. With no sample, $\sim 75$ V amplitude drive on the rf electrode results in secular-motion frequencies of $1.7$ MHz, $4.7$ MHz, and $6.4$ MHz for the $x$, $y$, and $z$ modes of a $^{25}$Mg$^+$ ion, respectively. A $1$ mT magnetic field is oriented at $25^{\circ}$ relative to the $X$-axis and $86.5^{\circ}$ with respect to the $Z$-axis [see Fig. 1(b)]. The magnetic field is used to lift the degeneracies of hyperfine levels and to define a quantization axis that provides well-defined beam propagation directions and light polarizations for driving cycling transitions and optical pumping. Doppler cooling is used to cool the $x$ and $y$ secular modes to a mean thermal occupation $\bar{n} \approx 7$, and Raman sideband cooling is used to ground-state cool below $\bar{n} < 0.1$ \cite{Monroe1995}. The $z$ mode is cooled weakly by Doppler laser beams and is not addressed by the Raman beams. Heating-rate measurements were conducted on the $y$ mode using Raman sideband spectroscopy (as described in Sec. II.C). To improve the sensitivity to electric-field noise from samples, the electric-field noise from the stylus trap was reduced by using $in$ $situ$ ion bombardment (Ar$^+$, 2 keV, $\sim1$ J/cm$^2$) \cite{Hite2012}. This results in a heating rate of $\sim 39$ quanta per second at $4.7$ MHz and $d_{t}=$ \SI{59}{\um}, which corresponds to an electric-field noise power spectral density at the ion position of $7 \times 10^{-13}$ V$^2$m$^{-2}$Hz$^{-1}$. 

\subsection{Sample description}
The \SI{4}{mm} $\times$ \SI{1}{mm} samples consist of electrodes patterned from \SI{5}{\um}-thick electroplated Au in either a meander or an interdigitated capacitor (IDC) geometry [see Fig. 1(c)]. On each sample, one electrode (green) is electrically grounded, while the remaining "signal" electrode (orange) can have electrical potentials applied to it [see Fig. 1(c)]. The samples were fabricated on the same quartz wafer and postprocessed under the same conditions; therefore, it is expected that the surface conditions of the two samples are similar. The trace width in the region of interest is \SI{6}{\um} with \SI{6}{\um} gaps. This geometry is designed to produce an exponential decrease in the electric-field strength with distance from the sample when potentials are applied to the signal electrode. The field resulting from a potential applied to the signal electrode will decay exponentially with distance normal to the surface with a characteristic length scale of $\sim$\SI{4}{\um}. This allows large potentials to be applied to the sample with minimal electric-field changes at the position of the ion along the $y$ mode. In this way, the dependence of surface electric-field noise on potentials applied to the sample electrode can be investigated. Further details on the electric field from applied potentials on the sample signal electrodes are provided in Appendix \ref{fieldappendix}.  The samples are wirebonded to a small circuit board, connecting the signal and grounded electrodes of each sample to the center pin and outer shield, respectively, of a coaxial cable. Each cable has a length of $\sim$\SI{240}{mm} in vacuum and is connected to a vacuum feedthrough. The signal electrodes on the sample can be terminated or biased by connecting appropriate circuity to the coaxial cable outside the vacuum feedthrough. The termination is typically a \SI{50}{\ohm} connection to ground unless described otherwise. 

The sample position is controlled using an $XYZ\Theta$ manipulator that allows for sub-micrometer step adjustments. The samples are mounted such that $\Theta$ rotations of the manipulator can select which sample is positioned into proximity with the trapped ion sensor. The initial sample height $h_0$ is calibrated by scanning the vertical position of a focused laser beam and measuring the obscuration of the transmitted beam by the trap features and sample chip edges. Changes in the height of the sample $\Delta h$ are measured using a high-precision digital contact sensor with \SI{0.1}{\um} resolution. The ion-to-sample distance $d_s$ is then given by $d_s=h-d_t=h_0-\Delta h - d_t$. The meander and IDC samples are tilted (unintentionally) relative to the stylus trap plane at angles of $\sim 0.5\pm0.5^{\circ}$ and $\sim 2.5\pm1.5^{\circ}$, respectively. This angular misalignment, combined with lateral positioning misalignments, results in a combined uncertainty in $h_{0,meander}$ of \SI{7.5}{\um} and $h_{0,IDC}$ of \SI{20}{\um}. 
\subsection{Electric-Field-Noise Measurement}
Previous measurements of electric-field noise from surfaces have shown strong dependence on distance \cite{Turchette2000, Deslauriers2006, Hite2017, Boldin2018, Sedlacek2018}, frequency \cite{Turchette2000, Deslauriers2006, Labaziewicz2008, Allcock2011, Hite2012, Bruzewicz2015}, and temperature \cite{Labaziewicz2008, Chiaverini2014, Sedlacek2018, Bruzewicz2015}. All of the measurements collected in this work were conducted at room temperature and at a secular frequency of $4.7$ MHz; therefore, in the remainder of this work, only the dependence on distance will be considered. The secular frequency was kept fixed for different values of $h$ by adjusting the amplitude of the rf drive. For the measurement setup presented in Fig. 1(d), the total electric-field noise power spectral density $S_{E,tot}$ measured can be modeled by 
\begin{equation}
    S_{E,tot} = S_{E,t}+S_{E,s},
\end{equation}
\begin{equation}
    S_{0} \equiv S_{E}\quad  \textrm{at}\quad  d=\SI{59}{\um},
\end{equation}
\begin{equation}
    S_{E,tot} = S_{0,t} (d_{t}/\SI{59}{\um})^{-\alpha_{t}}+S_{0,s} (d_{s}/\SI{59}{\um})^{-\alpha_{s}},
\end{equation}
\begin{equation}
    \dot{\bar{n}}_0 = \frac{q^2 S_{0}}{4m\hbar\omega},
\end{equation}
\begin{equation}
		\dot{\bar{n}}_{tot} = \dot{\bar{n}}_{0,t}\;(d_{t}/ \SI{59}{\um})^{-\alpha_{t}}+\dot{\bar{n}}_{0,s}\;(d_{s}/\SI{59}{\um})^{-\alpha_{s}},
\end{equation}
where the addition of subscripts $tot$, $t$, and $s$ represents total, trap, and sample, respectively. $S_0$ represents the electric-field noise power spectral density at a distance of \SI{59}{\um} (chosen because $d_{t0}=$ \SI{59}{\um}), $\dot{\bar{n}}$ is the heating rate in quanta per second, $\alpha$ represents the distance scaling exponent, $m$ and $q$ are the mass and charge of the ion, $\hbar$ is Planck's constant divided by $2\pi$, and $\omega$ is the angular frequency of the secular motion. In Ref. \cite{Hite2017}, a similar stylus trap (similar geometry and surface treatment) was used and $\alpha_t$ was determined to be $3.1$. We would expect the trap used in this work to exhibit similar characteristics. In this work, we conduct heating-rate measurements as a function of distance to extract values for $\dot{\bar{n}}_{0,t}$, $\dot{\bar{n}}_{0,s}$, $\alpha_t$, and $\alpha_s$. Heating-rate measurements are conducted by first cooling the $x$ and $y$ secular modes near their ground state ($\bar{n}<0.1$ ). After a variable delay time $t_{delay}$, the ratio of the red and blue Raman sideband amplitudes $r$, is used to determine the mean motional occupation $\bar{n}=r/(1-r)$. The heating rate is determined as the slope of a linear fit of $\bar{n}$ vs. $t_{\text{delay}}$. (For a brief description of heating-rate measurements see Ref. \cite{Hite2013}; for a detailed description see Ref. \cite{Wineland1998}). Prior to conducting heating-rate measurements with proximal samples, a number of experiments were conducted to confirm that technical noise sources that could contribute to ion heating were minimized or excluded. See Appendix A for a detailed discussion of the technical noise sources considered and the experimental checks conducted to characterize them. 

\section{Measurement Results}
Heating-rate measurements as a function of distance were conducted for the meander and IDC samples, where the signal electrode of each sample was connected to ground via a \SI{50}{\ohm} termination. Figure 2 shows the measured heating rates of the $y$ motional mode as a function of $d_{t}$ for the meander and IDC sample. 
\begin{figure*}[]
\centering
\includegraphics[width=0.9 \textwidth, keepaspectratio]{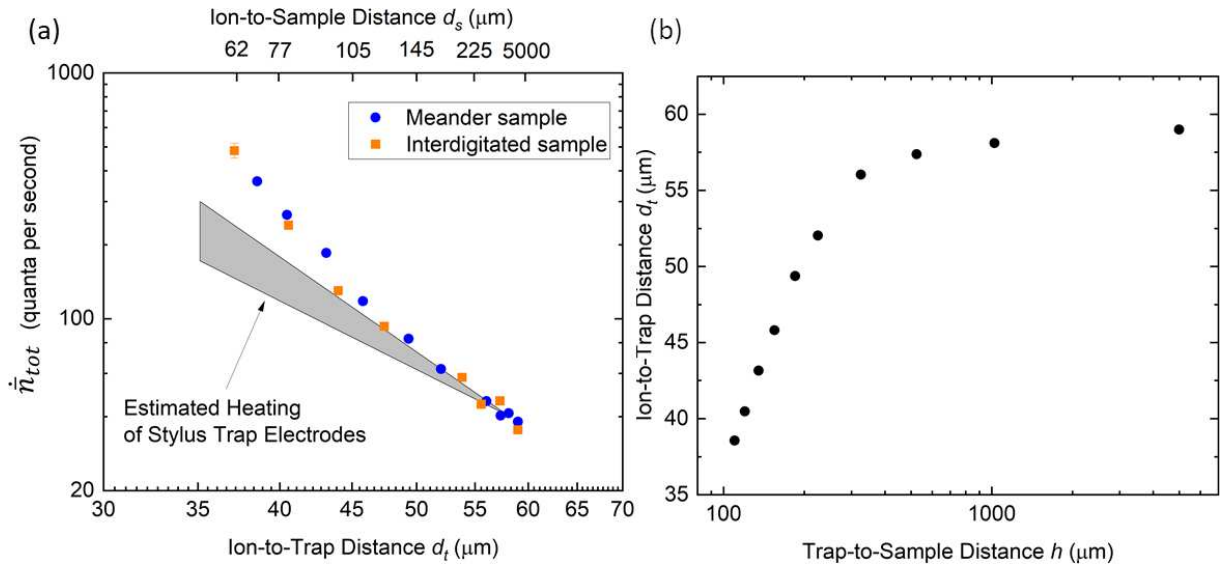}
\caption{a) Total heating rate $\dot{\bar{n}}_{tot}$ of the $y$ motional mode for the meander and IDC samples vs. ion-to-trap distance $d_t$. Shaded region represents expected heating rate due to the stylus trap alone, which is bounded by $d^{-3}$ scaling and $d^{-4}$ scaling. Error bars are smaller than the markers if not shown. Details on how uncertainties were derived are presented in Appendix C. The top axis scale shows measured values of $d_s$ for the meander sample data. b) Plot showing the relationship of ion-to-trap distance $d_t$ and trap-to-sample distance $h$. As the sample is positioned closer to the trap, the rf pseudopotential minimum moves closer to the trap. Note: Uncertainty in values of  $d_{t,0}=\pm$\SI{2}{\um}, $h_{0,meander}=\pm$\SI{7.5}{\um}, and $h_{0,IDC}=\pm$\SI{20}{\um} are not shown in either plot. Errors in the reported values due to the these uncertainties would result in a common systematic horizontal shift on all points, or a common systematic vertical shift as well in panel (b)   }
\label{fig:Data}
\end{figure*}
As the sample height $h$ is reduced from \SI{5}{mm} to \SI{110}{\um}, $d_t$ is reduced from its initial value of \SI{59}{\um} to \SI{39}{\um}, as shown in Fig. 2(b). The shaded region in Fig. 2(a) shows the expected heating background of the stylus electrodes, bounded by curves predicted from previous measurements of the distance scaling of ion heating rates with electrode distance ($3.1 < \alpha < 4$, see Table 1). 
\begin{table*}[]
    \centering
    \setlength{\tabcolsep}{18pt}
    \begin{tabular}{||c|c c c ||}
        \hline
        Trap Type & Scaling Exponent & Distance Range & Reference \\ [0.5ex] 
         \hline\hline
         Needle  & $3.5 \pm \; 0.1$ & $\sim40$ - \SI{200}{\um} & \cite{Deslauriers2006} \\ 
        \hline
         2D surface electrode & $3.79 \pm \; 0.12$ & 61 - \SI{154}{\um} & \cite{Boldin2018} \\
         \hline
        2D surface electrode & $4.0 \pm \; 0.2$ & 29 - \SI{83}{\um} & \cite{Sedlacek2018} \\
         \hline
        Stylus & $\sim 3.1$ & 34 - \SI{63}{\um} & \cite{Hite2017} \\
        \hline
        Stylus & $2.9 \pm \; 0.5$ & 38 - \SI{59}{\um} & this work \\ 
        \hline
        2D surface electrode & $3.9 \pm \; 0.5$ & 71 - \SI{1000}{\um} & this work \\ [1ex] 
        \hline
    \end{tabular}
    \caption{Summary of previous measurements for the distance scaling of ion heating rates with distance to nearest electrode. }
    \label{tab:my_label}
\end{table*}
The measured heating rates begin to increase above the predicted range of the trap heating-rate background for $d_t <$ \SI{49}{\um}, which corresponds to values of $d_s < $ \SI{135}{\um}. A fit to the meander sample data using Eq. 5 results in scaling exponents of $\alpha_t=2.9 \pm 0.5$  and $\alpha_s=3.9 \pm 0.5$ and is shown in Fig. 3(a) along with weighted residuals shown in Fig. 3(b). 
\begin{figure*}
\centering
\includegraphics[width=0.9 \textwidth, keepaspectratio]{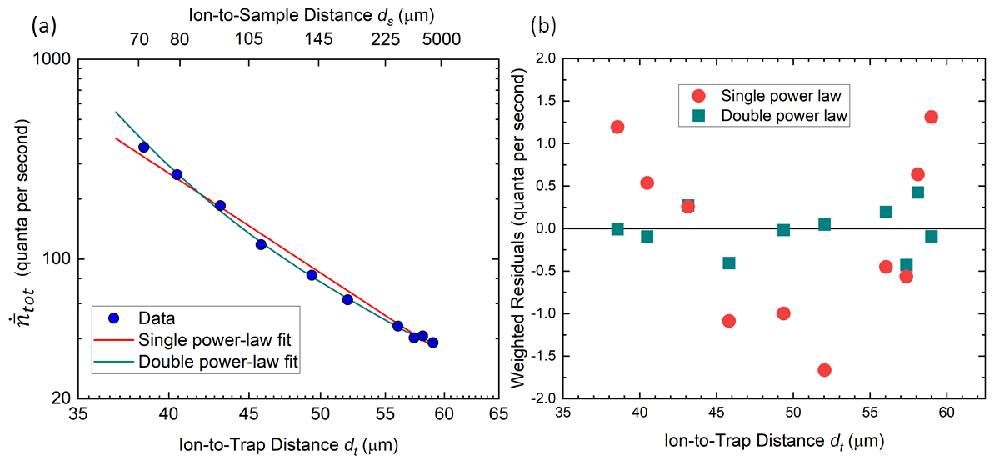}
\caption{a) Total heating rate $\dot{\bar{n}}_{tot}$ of the $y$ motional mode vs. ion-to-trap distance $d_t$, for the meander sample. The total heating rate $\dot{\bar{n}}_{tot}$ is fit to the sum of two power laws described in Eq. 5 as well as to a single power law. Error bars are smaller than the markers if not shown. Details on how uncertainties were derived are presented in Appendix C. Note: The double power-law fit line is an approximate visualization of the fit since the relationship between $d_t$ and $d_s$ is approximated by a fit function. For more details, see Appendix B. The top axis scale shows measured values of $d_s$ for the meander sample data. b) Residuals for fits shown in part (a). The residuals for the single power-law fit show systematic ``u''-shaped deviation from zero, while the residuals from the double power-law fit show a distribution more consistent with statistical noise. The sum of squares of the residuals (SSR) for the double power-law fit is $0.66$. For the single power-law fit, the SSR$ = 9.37$. Note: Uncertainty in values of  $d_{t,0}=\pm$\SI{2}{\um} and $h_{0,meander}=\pm$\SI{7.5}{\um} are not shown in either plot. Errors in the reported values due to the these uncertainties would result in a common systematic horizontal shift on all points. }
\label{fig:HRData}
\end{figure*}
Details on the uncertainties in the measured parameters and in the resulting fit can be found in Appendix C. 

The fitted value of $\alpha_t=2.9 \pm 0.5$ is consistent with the previous measurement of $\alpha_t=3.1$ for a similar geometry trap \cite{Hite2017}. The fitted value of $\alpha_s=3.9 \pm 0.5$ is consistent with the scaling for similar two-dimensional (2D) planar electrode surfaces \cite{Boldin2018, Sedlacek2018} (see Table 1). The fitted heating rate for the samples ($\dot{\bar{n}}_{0,s} = 500 {\tiny^\dscript{+300}{-200}}$ quanta per second at $d_s=$ \SI{59}{\um} for the meander sample) is within the range of reported heating rates in other measurements of room-temperature untreated electroplated Au traps \cite{Brownnutt2015}. 

The residuals for the double power-law fit suggest good agreement between the data and the model, but confirmation of the model requires an independent method for verifying at least one of the scaling exponents $\alpha_t$ and $\alpha_s$ or the sample-heating-rate scaling parameter $\dot{\bar{n}}_{0,s}$. In this particular experiment, it was not possible to independently measure any of these parameters. In future experiments, $d_t$ could be independently controlled with an rf potential applied to the center electrode, allowing for an independent measurement of $\alpha_t$ or to fix the value of $d_t$ independent of sample position, allowing for direct measurements of $\alpha_s$ or $\dot{\bar{n}}_{0,s}$. Instead, the double power-law fit results were compared to other potential models, such as a single power-law fit (first term only in Eq. 5) or a single power-law fit where the value of $\alpha_t$ changes with $d_t$ \cite{Turchette2000, Low2011}. The best fit to a single power law yields parameters of  $\alpha_t=5.2 \pm 0.2$ and $\dot{\bar{n}}_{0,t}=36 {\tiny^\dscript{+7}{-6}}$ for the meander sample data is shown in Fig. 3(a). The distribution of the weighted residuals for the single power-law fit shown in Fig. 3(b) display systematic effects and generally larger magnitude in comparison to those for the double-power-law fit, suggesting that the double-power-law model better describes the data. Additionally, the single-power-law exponent $\alpha_t=5.15$ is considerably higher than has been seen in other traps (see Table 1). One could alternatively fit the data to a single power-law model where the scaling parameter $\alpha_t$ is a function of $d_t$. A fit using this type of model would result in $\alpha_t=3$ for values of $d_t>$ \SI{55}{\um}, changing to $\alpha_t=6.5$ for values of $d_t<$ \SI{40}{\um}. This change in $\alpha_t$ is the opposite of the proposed behavior in Ref. \cite{Low2011}, where $\alpha_t$ is expected to get smaller with decreasing values of $d_t$. The change in $\alpha_t$ from 3 to 6.5 would also only be expected to occur for much larger variation in $d_t$, based on Ref. [20].

\section{Discussion}
In Ref. \cite{Hite2017}, an experiment similar to the one described in this work was conducted, but the behavior was different from what was observed in this work. In that work, Au samples were positioned at similar distances to the trap as the samples in this work, but the heating was consistent with a single power law with scaling exponent of $\alpha_t=3.1$. The conclusion of that work was that heating was due to electric-field noise from the trap electrodes and that electric-field noise from the samples was not detectable. In this work, the heating is consistent with the sum of power-law dependent terms for the trap and sample (Eq. 5). The reason for these different behaviors is unclear, but we performed tests to check possible hypotheses. 

Known differences between Ref. \cite{Hite2017} and this work are the sample electrode geometry and the wiring used to electrically bias the samples. The samples in Ref. \cite{Hite2017} were connected to a vacuum feedthrough using a \SI{0.26}{mm}-diameter wire approximately \SI{1}{m} in length. There was only one electrical connection per sample, and no samples had local ground connections; all biasing or grounding of a sample was via this wire. This wire was wound around a rod for strain relief purposes, resulting in several microhenries of inductance and a few kiloohms impedance at the rf drive frequency of 64 MHz. In contrast, the samples in this work were directly connected to a \SI{50}{\ohm} coaxial cable (see Sec. II.B) and terminated with a \SI{50}{\ohm} termination to ground. We tested for any dependence of measured heating rate on the impedance to ground seen by the sample signal electrode at the rf drive frequency. Variations in this impedance could affect the amplitude of induced currents or voltages in the sample from the rf trapping fields, which might potentially increase electric field noise from the sample surface by activating the motion of surface adsorbates, as postulated in Ref.~\cite{Hite2017}.  We positioned the IDC and meander samples at $h\sim$\SI{110}{\um} and used a $\uplambda/4$ resonator, made by connecting an appropriate length of coaxial cable with a short-circuit termination to ground to the vacuum feedthrough, to increase the impedance between the signal electrode and ground at the trap rf frequency.  Based on the estimated quality factor of the resulting resonance, which is set by cable loss, the impedance seen by the sample on resonance is approximately $3\pm1\,$\SI{}{\kohm}.  This increased rf impedance between the signal electrode and ground had no impact on the measured heating rates for either sample. We also swept the resonant frequency of the coaxial line resonator coupled to the sample by roughly $\pm$\SI{10}{\MHz} around the trap rf frequency using a coaxial line stretcher, which varies the magnitude of the load impedance at the trap rf frequency seen from the sample by roughly an order of magnitude. We observed no impact from these resonant frequency changes on the measured heating rates. We hope to continue to explore the impact of the impedance to ground for ion-trap electrodes on electric-field noise in future work. 

We also applied rf potentials to the signal electrodes of the IDC and meander samples to test for changes in the measured heating. A hypothesis that heating could be dependent on applied rf potentials was presented in Ref. \cite{Hite2017} and mentioned before that in Ref. \cite{Turchette2000} (Note: This should not be confused with RF-drive-related technical noise sources described in Ref. \cite{Wineland1998}). To test for any dependence on applied rf potentials, the signal electrodes of the IDC and meander samples were driven directly via their coaxial lines at 200 MHz, a frequency that is not resonant with any relevant ion transition frequencies. The IDC and meander samples were positioned at $h\sim$\SI{110}{\um} and pulses with shaped rising and falling edges to reduce any off-resonant excitations were applied to the samples during the delay time between ground-state cooling and sideband thermometry. Drive powers between $0$ and $200$ mW, corresponding to rf voltages on the electrodes with amplitudes of up to $\sim$\SI{9}{V}, were applied, but no significant dependence of the measured heating rate on the applied rf power was observed. 

\section{Conclusions}
In this work, we measured electric-field noise from interchangeable samples with an ion held in a stylus trap. The measured electric-field noise exhibited scaling with the ion-to-sample and ion-to-trap distances consistent with independent power-law distance dependencies for each surface. The power-law exponent, giving the distance scaling, as well as the multiplicative scaling factor, giving the absolute heating rate, were extracted for both the ion trap and sample surfaces. 

In the future, our trapped-ion-sensor system could be deployed in a vacuum chamber as part of a suite of precision surface-science measurement tools, such that test samples could be studied using multiple surface-characterization tools without breaking vacuum. In this way, it may be possible to perform high-throughput measurements of electric-field noise from test samples and to correlate the results with specific surface treatments or with data from other tools on surface characteristics such as morphology or chemical composition. 

To increase the utility of our technique as part of an integrated surface-characterization system as described above, we must reduce the uncertainty in the ion-sample and ion-trap distances, along with the overall electric-field noise due to the trap. Such improvements offer promise for understanding the origin of anomalous heating in ion traps, as well as for testing the performance of materials, treatment methods, and designs aimed at reducing anomalous heating. This could address one of the major outstanding challenges for large-scale trapped-ion quantum computing. 

\section{Acknowledgments}
The authors acknowledge helpful discussions with S. Glancy and E. Knill and thank M. Kim, A. McFadden, and R. Goldfarb for helpful comments on the manuscript. K.S.M. and P.D.K acknowledge support as associates in the Professional Research Experience Program (PREP) operated jointly by NIST and the University of Colorado Boulder under Award No. 70NANB18H006 from the US Department of Commerce, NIST. Contributions to this article by workers at NIST, an agency of the US Government, are not subject to US copyright. 

\appendix
\section{Technical Noise}
When interpreting heating-rate results, it is important to understand sources of technical noise in the experiment and to quantify effects that technical noise could have on the results. The sources of technical noise considered in our setup include Johnson noise from the electrodes and associated low-pass filter elements, noise from the digital-to-analog converters (DACs) and amplifiers used to generate dc potentials on the trap electrodes, environmental noise pickup, micromotion dependent noise \cite{Berkeland1998}, and stray resonant light. The calculated Johnson noise contribution to the electric-field noise at the ion position is more than two orders of magnitude smaller than the total measured electric-field noise. The DACs used in this experiment are heavily filtered to reduce noise at the secular frequencies and were tested by replacing the DACs with static voltages generated by resistive division of the potential from a battery. This configuration offers lower voltage noise than the DACs. There was no change in the measured heating rate when using these battery sources instead of the DAC voltage sources. Pickup on trap electrodes was reduced by eliminating sources of noise at the secular frequencies by measuring the radiation of various lab electronics. In-vacuum low-pass filters on the dc electrodes are located within a few centimeters of the trap to further reduce noise at the secular frequencies of the ion's motional modes. The rf electrode is bandpass-filtered by the helical resonator, and the geometry of the rf electrode means that voltages applied to the rf electrode do not produce electric fields at the pseudopotential null. If the ion position is offset from the rf null, micromotion is induced, which allows coupling to noise on the rf electrode at frequencies offset from the rf drive ($64$ MHz $\pm$ secular frequencies). The heating rate due to this source of noise would be proportional to the micromotion amplitude. By measuring heating rates as a function of micromotion amplitude and observing no dependence, this noise source can be shown to be insignificant. Stray resonant light can be absorbed and cause spontaneous emission during the delay time of the heating-rate measurement, resulting in recoil heating. Detectors with picowatt sensitivity were used to check for any stray resonant light present in the beam paths \cite{Arrington2013}. 

The electric field at the ion position from potentials on the signal electrode of the samples was designed to be minimized by the electrode meander or IDC geometry (see Sec. II.B and Appendix D). However, potentials on the wire-bonding pads, or the leads between those pads and the meander/IDC regions, as well as other effects described in Appendix D, could give rise to electric fields at the ion position; this effect would allow technical noise on the sample signal electrodes\textemdash which have no in-vacuum filtering\textemdash to cause heating of the ion. To understand the heating rate contribution from technical noise on the sample signal electrode, we performed a measurement of the relative electric field strength from the signal electrode at the ion position as a function of $h$. We applied an oscillating voltage pulse of fixed duration of the form $V_0(h)\sin{\omega_y t}$, resonant with the motional secular frequency, to the sample electrode and monitored the amplitude of the resulting coherent excitation of the ion motion. For each value of $h$, we adjusted the voltage amplitude $V_0(h)$ to produce the same total motional excitation of the ion (determined as the excitation level where the resonant fluorescence from the ion during state detection was reduced to half of its value without the applied voltage), indicating the same electric field amplitude at the ion. The values of $V_0(h)$ are thus inversely proportional to the electric field amplitude at the ion produced by a given voltage on the sample signal electrode, for each value of $h$.

We then consider the case of technical noise on the signal electrode, with voltage spectral density $S_V^{1/2}(\omega)$ that is independent of $h$. In this case, the resulting electric-field noise power spectral density at the motional frequency $S_{E, tech}(\omega_y)$, and thus the heating rate contribution from technical noise $\dot{\bar{n}}_{tech}$, will depend on $h$ as $[V_0(h)]^{-2}$. Figure 4 plots $[V_0(h)]^{-2}$ vs $d_t$ for the meander and IDC samples. 
\begin{figure}[ht!]
\centering
\includegraphics[width=8 cm, keepaspectratio]{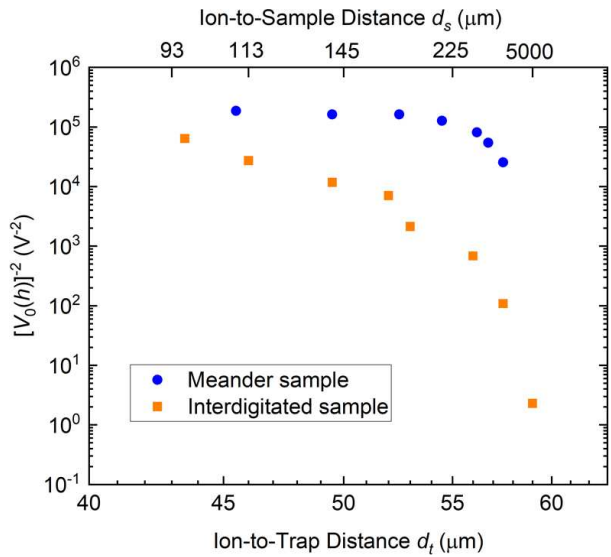}
\caption{Plot of $[V_0(h)]^{-2}$ vs. the ion-to-trap distance $d_t$. The expected heating rate contribution $\dot{\bar{n}}_{tech}$ from technical noise on the sample signal electrode is proportional to the plotted $[V_0(h)]^{-2}$. Both the distance scaling and the magnitude of the expected $\dot{\bar{n}}_{tech}$ are different between the meander and the IDC samples, likely due to the differences in bond pad and trace geometry of the two samples [see Fig. \ref{fig:Trap}(c)] or to differences in the electric fields arising from phase gradients along the sample signal electrodes as described in Eq.~\ref{eq:gradientv}. The top $x$-axis label represents measured values of $d_s$ for the meander sample data. Uncertainty in values of  $d_{t,0}=\pm$\SI{2}{\um}, $h_{0,meander}=\pm$\SI{7.5}{\um}, and $h_{0,IDC}=\pm$\SI{20}{\um} are not shown. Errors in the reported values due to the these uncertainties would result in a common systematic shift on all points.}
\label{fig:tickle}
\end{figure}

Figure 4 shows that $[V_0(h)]^{-2}$ scales very differently for the two samples, with neither sample exhibiting expected technical noise heating rates $\dot{\bar{n}}_{tech}$ that scale as $\sim d_s^{-4}$. Additionally, an equivalent magnitude of technical noise on the two samples would result in a technical heating rate about five times larger for the meander sample than for the IDC at $d_t\sim$ \SI{45}{\um}. Since the observed heating rates for the two samples are very similar in both magnitude and distance dependence [see Fig. 2(a)], we conclude that technical noise on the signal electrodes of the samples is not the dominant source of the observed heating rates. We provide further support for this conclusion by calculating the predicted scaling with $d_s$ of the electric field amplitude at the ion position from the sample signal electrode due to various geometric considerations and effects (see Appendix D) and showing that this scaling is different from that extracted from the double-power-law fit.

\section{Double Power-Law Fit Graphics}
The relationship between $d_t$ and $h$ shown in Fig. 2(b) was determined empirically from measurements at the plotted points. The double power-law fit shown in Fig. 3(a) [residuals in Fig. 3(b)] was conducted based on these empirically determined points. In order to plot a fit line for Eq. 5 in Fig. 3(a), it was necessary to interpolate values for $d_s=h-d_t$ for graphical purposes. While it might be possible to determine an analytical solution to describe $d_t$ vs. $h$, we instead used an asymptotic exponential Lorentzian function to approximate this relationship. The equation used to generate the relationship for $d_t$ vs. $h$ is 
\begin{equation}
    h\approx-40.2+\frac{3881*0.81}{4(d_t-59.05)^2+0.81^2}-197\;\text{log}(1-\frac{d_t}{73.46}) 
\end{equation}
(values rounded). The fit line shown used this function to approximate the interpolated values of $d_s$ and is therefore only a visual guide to the fit. The weighted residuals shown in Fig. 3(b) are an exact representation of the fit. 

\section{Measurement Uncertainties and Statistics}
This section describes how the reported  uncertainties were determined and their impact on the reported fit parameters. Section II.C describes how a heating rate (linear fit of $\bar{n}$ vs. $t_{\text{delay}}$) is measured. The linear heating-rate fit is weighted based on the standard error of the means of each value of $\bar{n}$. The result of the fit is a value for the heating rate and an uncertainty estimate on that heating rate. The heating-rate measurements are repeated, and then a weighted mean and weighted uncertainty for $\dot{\bar{n}}$ (weighted using the uncertainties for each linear heating rate fit) is calculated. These weighted values are plotted in Fig. 2 and Fig. 3. The uncertainties in the values of $h_0$, $d_t$, and $d_s$ were discussed in Sec. II.
To quantify the impact of the uncertainties, bootstrapping \cite{Efron1981, Efron1987} was used to generate 5000 simulated datasets,  where $\dot{\bar{n}}$, $d_{t0}$, and $h_{0}$ were drawn at random from Gaussian distributions with means and standard deviations given by the estimates and uncertainties, respectively, of the calibrated values. The generated datasets were then fit to Eq. 5, and the resulting distributions in fit parameters were used to generate 68\% bias corrected confidence intervals for the fit parameters. The resulting asymmetric $68\%$ confidence intervals were then rounded to an appropriate significance level and reported accordingly in the main text.  

The meander dataset consisted of measurements conducted at ten values of $d_t$ and $d_s$. Fitting to Eq. 5 results in five degrees of freedom, which is scarce for multiple power-law dependencies. In future measurements of the distance dependence, we plan on measuring heating rates at a larger number of distance values to increase the number of degrees of freedom represented by the data. The IDC dataset only contained eight points, leaving just three degrees of freedom. The limited degrees of freedom, combined with much larger uncertainty in $h_{0}$, resulted in a fit with very large uncertainty values. The distance dependence data for the IDC dataset was conducted during commissioning of the experiment, which also added additional uncertainty to the results. The complete distance measurement was not repeated prior to disassembling the experiment, and as a result, the confidence level in the measurements of $d_s$ for the IDC data is low. However, for completeness, the measured heating rates as a function of $d_t$ were still shown in Fig. 2(a).

\section{Electric field from sample electrodes\label{fieldappendix}}

This section discusses the electric field at the ion position due to applied potentials on the signal electrodes of the samples.  We first present a derivation of the exponential decay with distance from the electrode plane of the electric field parallel to the electrode plane.  We then consider the effects of imperfect centering of the sample over the ion, or sample tilt with respect to the $(x,y)$ plane of the ion motional modes, on the electric field along the $y$ motional mode due to potentials applied to the sample signal electrode.  Finally, we consider impacts related to the non-zero driving frequency of the applied potentials.  

We define a coordinate system $(\tilde{x},\,\tilde{y},\,\tilde{z})$ relative to the sample electrodes, where $\tilde{x}$ is in the plane of the sample electrodes but perpendicular to the long axis of the meander/IDC segments, $\tilde{y}$ is along the length of the meander/IDC segments, and $\tilde{z}$ is normal to the surface, pointing away from the substrate.  To begin, we assume that the sample electrode has infinite extent in the $(\tilde{x},\,\tilde{y})$ plane (we will discuss finite-size corrections later).  

To solve for the potential $\phi(\tilde{x},\,\tilde{y},\,\tilde{z})$ in the half-space above the electrodes ($\tilde{z}>0$), we first recognize that there is continuous translational symmetry along $\tilde{y}$, and therefore $\phi$ must be independent of $\tilde{y}$.  There is discrete translational symmetry along $\tilde{x}$ with periodicity {$a=24$\,\si{\micro\metre}} from the repeating pattern of alternating ground and signal electrodes, which allows us to express the $\tilde{x}$-dependence of the potential by Fourier decomposition.  We can therefore make an ansatz for the potential $\phi$ of the form 
\begin{equation}
\label{phiansatz}
    \phi(\tilde{x},\,\tilde{z})=\sum_{n=0}^\infty A_n(\tilde{z})\cos{\frac{2\pi n\tilde{x}}{a}}.
\end{equation}
We can choose the location of the origin $\tilde{x}=0$ without loss of generality, and thus can use cosines as the Fourier basis functions.  We then solve the Laplace equation $\nabla^2\phi=0$, which here reduces to $\frac{\partial^2 \phi}{\partial \tilde{z}^2} = -\frac{\partial^2 \phi}{\partial \tilde{x}^2}$, to determine the functional form of the Fourier coefficients $A_n(\tilde{z})$, giving
\begin{equation}
    \sum_{n=0}^\infty \frac{\partial^2 A_n(\tilde{z})}{\partial \tilde{z}^2}\cos{\frac{2\pi n\tilde{x}}{a}}=\sum_{n=0}^\infty \frac{4\pi^2 n^2 A_n(\tilde{z})}{a^2}\cos{\frac{2\pi n\tilde{x}}{a}}.
\end{equation}
The orthogonality of the cosines in the sums requires this equality to hold for each value of $n$ individually, so that we find
\begin{equation}
    \frac{\partial^2 A_n(\tilde{z})}{\partial \tilde{z}^2}=\frac{4\pi^2 n^2 A_n(\tilde{z})}{a^2}.
\end{equation}
 When $n=0$, the solution is $A_0(\tilde{z}) = k_0 \tilde{z} + k_1$, where $k_0$ and $k_1$ are constants.  For $n\neq0$, the solution to this equation is 
\begin{equation}
A_n(\tilde{z}) = C_n\, e^{-2\pi n \tilde{z}/a} + D_n\, e^{2\pi n \tilde{z}/a},
\end{equation}
where $C_n$ and $D_n$ are constants. For $\tilde{z}>0$, $D_n$ must be zero because of the boundary condition that $\phi\rightarrow 0$ as $\tilde{z}\rightarrow\infty$.  

The total electric field $\nabla\phi$ can then be expressed in terms of its components as
\begin{align}
    E_{\tilde{x}} &= \frac{2\pi}{a}\sum_{n=1}^\infty n C_n\, e^{-2\pi n \tilde{z}/a}\cos{\frac{2\pi n\tilde{x}}{a}}\\
    E_{\tilde{z}} &= k_0 - \frac{2\pi}{a}\sum_{n=1}^\infty n C_n\, e^{-2\pi n \tilde{z}/a}\cos{\frac{2\pi n\tilde{x}}{a}}\, .
\end{align}
Thus we see that the electric field from the sample electrodes consists of a uniform field along $\tilde{z}$ plus other contributions in both $\tilde{x}$ and $\tilde{z}$ whose amplitude dies off exponentially with $\tilde{z}$ on a characteristic length scale given by $a/(2\pi n)$. Since this length scale is smaller for increasing $n$, the dominant contribution will come from the $n=1$ term that decays as $a/(2\pi)$, which is $\approx4$\si{\micro\metre} in our case.  Thus at distances from the sample electrode plane $\gg a/(2\pi)$, which is the case in our experiments, the electric field parallel to the sample electrode plane is negligible (subject to some caveats to be described next).  More generally, at distances from the surface $\gg a/(2\pi)$, the electric field from biasing the signal electrode at potential $V_0$ is well-approximated by treating the sample as a uniform conductor biased at potential $V_0/2$, the average potential of the signal and ground electrodes (which have equal areas).  In practice, if one approximates the plane of the stylus trap electrodes as a uniform ground that is parallel to the sample electrode surface and separated by $h$, then the magnitude of the electric field along $\tilde{z}$ will be $V_0/(2h)$.  

The real sample electrode has a finite extent $\ell\approx1\,$\si{mm} in $\tilde{x}$ and $\tilde{y}$, which must be accounted for.  Since we are always operating in the limit $\tilde{z}\gg a/(2\pi)$, we can use the approximation of the sample electrode as a uniform conductor biased at $V_0/2$.  We define the origin $(\tilde{x}=0,\tilde{y}=0,\tilde{z}=0)$ to be on the sample surface at the center of the sample electrode active area, which is $\ell\times\ell$.  Ideally, the ion position is at $(\tilde{x}=0,\tilde{y}=0,\tilde{z}=d_s)$ in this coordinate system, and the electric field will be purely along $\tilde{z}$ by symmetry.  However, if the ion is at nonzero values of $\tilde{x}$ and/or $\tilde{y}$, then there will be a nonzero electric field along the direction of displacement from $(0,0,d_s)$.  For an ion positioned at $(\tilde{x}=\varepsilon_{\tilde{x}} ,\tilde{y}=\varepsilon_{\tilde{y}},\tilde{z}=d_s)$, where $\{\varepsilon_{\tilde{x}},\varepsilon_{\tilde{y}}\}\ll\ell$ and $\{\ell,d_s\}\gg a/(2\pi)$, the electric field will be approximately
\begin{align}\label{eq:offsetv}
    E_{\tilde{x}}&\approx \sum_{j=-\infty}^\infty\frac{32 \ell (d_s+2 j h) V_0}{\left[\ell^2+4(d_s+2 j h)^2\right]^2}\varepsilon_{\tilde{x}} \\
    E_{\tilde{y}}&\approx \sum_{j=-\infty}^\infty\frac{32 \ell (d_s+2 j h) V_0}{\left[\ell^2+4(d_s+2 j h)^2\right]^2}\varepsilon_{\tilde{y}}\label{eq:offsetvmid}\\    
    E_{\tilde{z}}&\approx \frac{V_0}{2h}\, .\label{eq:offsetv2}
\end{align}

Since in our system the $(\tilde{x},\tilde{y})$ plane of the sample electrodes is parallel to the $(x,y)$ plane of the ion motional modes~\footnote{The $\tilde{y}$ axis is only approximately parallel to the $y$ axis, and the $\tilde{x}$ axis is only approximately parallel to the $x$ axis, although the two planes are parallel to a high degree.}, imperfect centering of the sample over the ion in the $\tilde{y}$ direction when a potential is applied to the sample signal electrode will give rise to an electric field contribution along the $y$ motional mode with strength linearly proportional to the offset from center.  The summation over $j$ captures the effect of the presence of the stylus trap, which we approximate as an infinite ground plane at a distance $h$ from the sample electrode plane~\cite{Schmied2011}.  The effect of this ground plane on the electric fields in $\tilde{x}$ and $\tilde{y}$ depends on $d_s$.  At distances $d_s\gtrsim\ell$, $E_{\tilde{x}}$ and $E_{\tilde{y}}$ can be roughly approximated by the respective $j=0$ terms in the sums in Eqs.~\ref{eq:offsetv} and \ref{eq:offsetvmid}, which correspond to the case in the absence of the ground plane. When $d_s\ll\ell$, the presence of the ground plane suppresses these electric fields by a factor of roughly $\exp{\ell/(2 d_s)}$ relative to the $j=0$ (no ground plane) case.    

If the $(\tilde{x},\tilde{y})$ plane of the sample electrodes is tilted slightly with respect to the $(x,y)$ plane of the ion motional modes, then the $E_{\tilde{z}}$ field can have a component along the ion's $y$ motional mode.  The strength of this field will depend inversely on $h$.  

A final consideration is specific to oscillating applied potentials on the sample signal electrode, which we take to be of the form $V_0\cos{\omega_d t}$.  Although the meander and IDC samples can both be treated approximately as lumped-element capacitive terminations on the end of a transmission line, both the meander line and the interdigitated fingers of the IDC have finite (non-zero) electrical length.  As a result, there is a small phase gradient in the applied voltage along the length of the sample signal electrodes.  Based on finite-element simulations, this phase gradient is proportional to $\omega_d$ and is $\sim 0.17$ at $\omega_d/2\pi=200$ MHz for both samples.  Since the phase difference between the ends of the sample signal electrodes is much less than $2\pi$, we can approximate the electrode potential as having a linear spatial gradient that oscillates in time, in addition to the spatially uniform potential.  By symmetry, this gradient creates no electric field along the $\tilde{z}$ direction, but gives rise to an electric field at the ion position in the $\tilde{x}$ direction (for the meander) or $\tilde{y}$ direction (for the IDC) of magnitude
\begin{align}\label{eq:gradientv}
    |\vec{E}|&\approx\frac{8\pi V_0 \omega_d}{\ell\,\omega_\lambda}\cos{\omega_d t} \nonumber\\
    &\times\sum_{j=-\infty}^\infty\left[\frac{2 \ell (d_s+2 j h)}{\ell^2+4(d_s+2 j h)^2}+\arctan{\frac{\ell}{2 (d_s+2 j h)}}\right]\, , 
\end{align}
where $\omega_\lambda/2\pi\approx7.2$ GHz (representing the frequency at which the sample signal electrodes are one wavelength long), and we have assumed that $\varepsilon_{\tilde{x}}=0$ and $\varepsilon_{\tilde{y}}=0$.  The magnitude of the field in Eq.~\ref{eq:gradientv} scales as $d_s^{-4}$ for $d_s\gg\ell$, and is approximately constant for $d_s\ll\ell$.  Here, the presence of the stylus trap (approximated as an infinite ground plane) does not provide substantial suppression of the electric field for $d_s\ll\ell$.  The non-spatially-varying portion of the potential will produce electric fields of the form of Eqs.~\ref{eq:offsetv}-\ref{eq:offsetv2} multiplied by $\cos{\omega_d t}$; whether these are larger or smaller than the fields from the spatial gradient in Eq.~\ref{eq:gradientv} depends on the drive frequency $\omega_d$ and the offsets from center $\{\varepsilon_{\tilde{x}},\varepsilon_{\tilde{y}}\}$, as well as any suppression in the $d_s\ll\ell$ regime as described above.


\end{document}